# Gigahertz-rate-switchable wavefront shaping through integration of metasurfaces with photonic integrated circuit

**Haozong Zhong,**[a, †] **Yong Zheng,**[a, †] **Jiacheng Sun,**[b, †] **Zhizhang Wang,**[b] **Rongbo Wu,**[a] **Ling-en Zhang,**[a] **Youting Liang,**[a] **Qinyi Hua,**[a] **Minghao Ning,**[a] **Jitao Ji,**[b] **Bin Fang,**[c] **Lin Li,**[a, e, *] **Tao Li,**[b, *] **Ya Cheng,**[a, d, e, *] **Shining Zhu**[b, *]

[a]State Key Laboratory of Precision Spectroscopy, School of Physics and Electronic Science, East China Normal University, Shanghai 200241, China
[b]National Laboratory of Solid State Microstructures, College of Engineering and Applied Sciences, Nanjing University, Nanjing, 210093, China
[c]College of Optical and Electronic Technology, China Jiliang University, Hangzhou 310018, China
[d]State Key Laboratory of High Field Laser Physics and CAS Center for Excellence in Ultra-Intense Laser Science, Shanghai Institute of Optics and Fine Mechanics (SIOM), Chinese Academy of Sciences (CAS), Shanghai 201800, China
[e]Collaborative Innovation Center of Extreme Optics, Shanxi University, Taiyuan 030006, China

*Lin Li, lli@lps.ecnu.edu.cn; Tao Li, taoli@nju.edu.cn; Ya Cheng, ya.cheng@siom.ac.cn; Shining Zhu, zhusn@nju.edu.cn

†These authors contributed equally to this work.

**Abstract**: Achieving spatiotemporal control of light at high-speeds presents immense possibilities for various applications in communication, computation, metrology, and sensing. The integration of subwavelength metasurfaces and optical waveguides offers a promising approach to manipulate light across multiple degrees of freedom at high-speed in compact photonic integrated circuit (PICs) devices. Here, we demonstrate a gigahertz-rate-switchable wavefront shaping by integrating metasurface, lithium niobite on insulator (LNOI) photonic waveguide and electrodes within a PIC device. As proofs of concept, we showcase the generation of a focus beam with reconfigurable arbitrary polarizations, switchable focusing with lateral focal positions and focal length, orbital angular momentum light beams (OAMs) as well as Bessel beams. Our measurements indicate modulation speeds of up to gigahertz rate. This integrated platform offers a versatile and efficient means of controlling light field at high-speed within a compact system, paving the way for potential applications in optical communication, computation, sensing, and imaging.

## Introduction:

The remarkable multiple degrees of freedom (DOFs) possessed by light, including large bandwidth and high-speed transmission capabilities, make photonic technology an extremely promising platform for high-speed communication and high-performance computing in information science[1-3]. Over the last decade, metasurfaces have emerged as an unprecedented approach for manipulating light in various DOFs using compact, artificial two-dimensional nanostructures[4,5]. Metasurfaces have demonstrated significant promise for a wide range of applications in both fundamental science and industry, including wavefront shaping[6,7], polarization control[8-10], imaging[11,12], spectrometry[13,14], computation[15]. Recently, considerable research efforts have been devoted to developing metasurfaces to achieve tunable or reconfigurable functionalities[3,16]. A high-speed spatiotemporally controlled metasurface possesses the potential to facilitate novel physics and practical applications within the realm of photonic technology. Numerous materials and strategies are proposed to empower tunable metasurfaces, including the utilization of phase change materials, liquid crystal, thermo-optic effect and electro-optic effect[17] [18-22]. However, effectively exploring tunability in multiple DOFs with high-speed and sufficient efficiency remains a significant challenge in metasurface research.

On the other hand, photonic integrated circuits (PICs) represent an alternative to traditional electronic technologies by utilizing light. Diverse applications have been demonstrated with PICs, including high-speed optical communication, signal processing, computing, and emerging technologies in quantum, biomedicine, and sensing[23-28]. Particularly, the advent of lithium niobate on insulator (LNOI) has propelled PICs as a promising platform for future high-speed electro-optic (EO) integrated devices[29], such as high-performance modulators[30,31], frequency combs[32,33], polarization controllers[34,35], and quantum optics circuits[36,37]. However, the full utilization of the DOFs of light in traditional two-dimensional PICs has not been realized, thereby limiting their application in optical information technologies. Recently, the integration of metasurfaces with waveguides has enabled PICs to manage multiple DOFs of light in both free space and waveguides with on-chip devices, providing a highly integrated platform for efficient photonics management[38]. A variety of applications have been reported with the guided-wave-driven metasurfaces, such as mode conversion[39,40], on-chip lenses[41,42], holography[43,44], OAM generator[3,45-48], as well as polarization control[49,50]. Nevertheless, achieving high-speed wavefront shaping and switching capabilities through the integration of metasurfaces, photonic waveguides, and electrical circuits remains a significant scientific and technological challenge for PICs and metasurfaces.

In this work, we propose and demonstrate a gigahertz-rate-switchable wavefront shaping by integrating metasurfaces with LNOI PICs. An arbitrary polarized light could be generated by combing a waveguide with two orthogonally propagated modes and nanoscatters at the specific positions[45]. Through the introduction of a Mach-Zehnder interferometer (MZI) and a phase shifter together with two pairs of electrodes, both the amplitude and the phase of the two orthogonal modes could be managed, enabling the generation of light with arbitrary polarizations spanning the entire surface of Poincaré sphere at high-speed. Meanwhile, a well-designed polarization dependent metasurface is introduced to achieve the desired functionality and facilitate high-speed modulation or switching. With this stratagem, switchable focusing with lateral focal positions and focal length, OAMs as well as Bessel beams are demonstrated. By effectively combining the propagation phase and geometric phase of birefringent nanostructures within this waveguide scheme, we demonstrate the switchability of these functionalities could be in arbitrary orthogonal polarizations. The switching speed reaches gigahertz rates, while the modulation speeds can be optimized to reach hundreds of gigahertz using tailored electrodes and LNOI PIC waveguides. This approach provides a versatile and efficient means of controlling light propagation in a compact and integrated system with simple electrical wiring and low power consumption, and promises important advantages in scenarios such as optical communication, computation, sensing, and imaging.

**Results and Discussion**

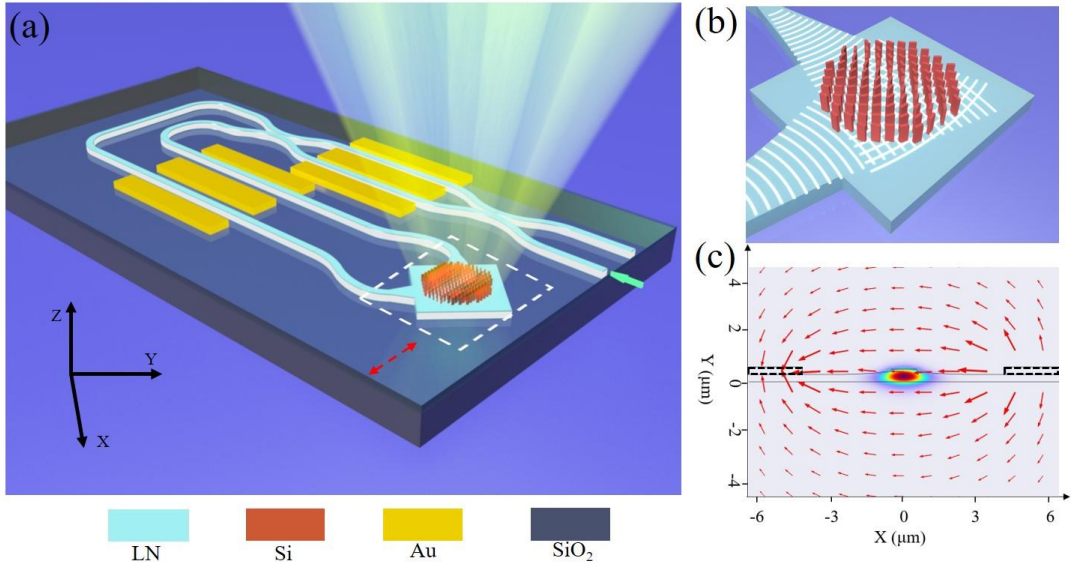

**Figure 1.** (a) Schematic diagram of PIC-driven metasurface device with LNOI, the red arrow is the direction of optical axis. (b) The zoomed-in scheme of the integrated metasurface, depicted with the fictitious wavefront of the waveguide modes. (c) The simulated static electric field when a 1 V voltage is applied between the electrodes, superimposed with the simulated optical field profile of the TE0 mode in the LN ridge waveguide. The black dashed lines indicate the electrodes. They have a height of 300 nm and are spaced apart by 6.5 μm. The top width of the LN ridge waveguide is 1μm, and the etch depth is 210 nm.

Figure 1(a) schematically show the PIC-driven metasurface device. Two fundamental transverse-electric ($TE_0$) modes from LN ridge waveguides are transform into two orthogonally propagated $TE_0$ modes in the slab waveguide. This transformation is achieved through the use of adiabatic tapers, as illustrated within the dashed square in Fig. 1(a). The zoomed-in view of the region is provided in Fig. 1(b). The electromagnetic wave in the slab waveguide can be expressed as:

$$E_{xy} = \begin{bmatrix} Ae^{i(\varphi_x + x_n \beta_x)} \\ Be^{i(\varphi_y + y_n \beta_y)} \end{bmatrix} \quad (1)$$

Where $A$, $B$, $\varphi_x$ and $\varphi_x$ represent the amplitudes and initial phase of the two guided waves, respectively. $\beta_x$, $\beta_y$ are the propagation constants of the guided waves in slab waveguide. The near orthogonal nature of the two transverse components of the electric field enables the synthesis of a wide range of polarization states across the entire surface of the Poincaré sphere. This synthesis is achieved by manipulating the amplitudes and phase lags ($\Delta\varphi = \varphi_x + x_n\beta_x - \varphi_y - y_n\beta_y$) of the two components. To enable high-speed modulation, an electrically controlled Mach-Zehnder interferometer (MZI) and a phase shifter are utilized to modulate the amplitude and phase of the two components. Subsequently, guided-wave-driven silicon metasurfaces are strategically positioned on the slab waveguide, with specific uniform polarizations, to generate variety electrically controlled functionalities. These functionalities can be manipulated at high speeds, offering enhanced control and versatility.

The device is designed and fabricated on a X-cut LNOI platform to provide the best electro-optical performance. Considering of the anisotropy of X-cut LN, both $TE_0$ modes are specifically designed to propagate at a 45° angle to the short optical axis of LN. This arrangement ensures that the wave-vectors $\beta_x=\beta_y$ and maintains a symmetric polarization distribution, as shown in Fig. 1(a). To precisely get the polarization distribution on the waveguide, we perform a finite-difference time domain (FDTD) simulation to get the amplitude and phase evolution over the slab waveguide at the experiment wavelength of 1550 nm as shown in the supplementary materials (SM) Fig. S1. Directional couplers (DC) are employed to achieve the tunable splitting function of the MZI, having a center-to-center spacing of 4.5μm and a length of 300μm. Meanwhile, ground-signal-ground (G-S-G) electrodes are used in the inner and outer phase shifters of MZI to induce phase shifts in a single-drive push–pull configuration, so that the electric field induce phase shifts in both arms with equal magnitude but opposite sign. The simulated optical field profile of the $TE_0$ mode in the LN ridge waveguide together with the static electric field between the electrodes are shown in Fig. 1(c), exhibiting well overlap of the two fields.

The PIC-driven metasurface is initially investigated to showcase a functionality with arbitrary reconfigurable polarization. In order to demonstrate the feasibility of this approach, a focusing beam achieved with silicon nano-cylinder metasurface is investigated. The Jones matrix of isotropic silicon nano-cylinders are given by $e^{i\alpha}\begin{pmatrix}1 & 0\\ 0 & 1\end{pmatrix}$, where $\alpha$ is the propagation phase of the nano-cylinders. The electromagnetic wave extracted to free space by silicon cylinders can be expressed as:

$$\begin{bmatrix}E_{xout}\\ E_{yout}\end{bmatrix}=e^{i\alpha}\begin{bmatrix}1 & 0\\ 0 & 1\end{bmatrix}\begin{bmatrix}Ae^{i(\varphi_x+x_n\beta_x)}\\ Be^{i(\varphi_y+y_n\beta_y)}\end{bmatrix}=e^{i\alpha}\begin{bmatrix}Ae^{i(\varphi_x+x_n\beta_x)}\\ Be^{i(\varphi_y+y_n\beta_y)}\end{bmatrix} \quad (2)$$

Thus, the phase profile of free space electromagnetic wave is depended on location of the silicon cylinders. On the other hand, to generate a focused beam in free space, the phase of light scattered from the metasurface has a distribution described below

$$\varphi(x,y)=-\frac{2\pi}{\lambda}(\sqrt{x^2+y^2+f^2}-f) \quad (3)$$

where $f$ is the distance between the device plane and the focal plane, that is, the focal length. Then, the metasurface is constructed by arranging this set of silicon cylinders on top of the slab waveguide to extracted the desired polaritons state with the phase profile for focusing. Here, the focal length is designed to be $f= 50\ \mu m$. The silicon cylinders have the identical dimensions, with radius of R=100 nm and height of H=1000 nm. The detailed design process of the metasurface array and the fabrication process are described in Supplementary Material Sections 2 and 7, respectively. Figure 2(a) displays the microscope image of the fabricated device, where the electrodes have a length of 5 mm and a gap of 6.5 μm between them. Figure 2(b) shows the scanning electron microscopy (SEM) image of the fabricated metasurface. We captured the three-dimensional (3D) scattered light field distribution above the device using a home-built optical setup (SM section 6). An image recorded at *X-Z* plane is illustrated in Fig. 2(c), revealing a distinct focal spot with a full-width-half-maximum (FWHM) of approximately 2 μm.

By adjusting the voltages applied to the electrodes of the MZI and the phase shifter, the polarization state of focal spot can be dynamically manipulated. To evaluate the performance of the reconfiguration polarizations of our device, two triangle wave signals with frequencies of 1 KHz and 100 KHz are applied to the electrodes of the MZI and phase shifter, respectively. The peak-to-peak drive voltages (Vpp) are set to 9.2 V, corresponding to a phase shift of approximately $2\pi$. The blue dots in Figure 2(d) depicts the measured sampling points on the Poincaré sphere (The schematic diagram of the experimental setup is presented in SM, section 6 and the corresponding measured Stokes parameters is presented in SM, section 7), representing the polarization states of the focal spot. The polarization extinction ratio (PER) was measured to be approximately 20 dB (SM section 3). The majority of the Poincaré sphere's surface is covered, with the exception of two areas near the north and south poles. This deviation is mainly due to imperfections in the fabrication process, resulting in a splitting ratio of the directional coupler (DC) in the MZI that deviates from the ideal 50:50 ratio. This issue can be addressed by employing an additional electrically controlled interferometer to improve the splitting ratio[34]. Figure 2(e) depicts the well stability of the generated polarization state, indicating its robustness. Figure 2(f) showcases the switching performance between two polarizations, highlighting the high repeatability of polarization generation. By adjusting the voltages applied to the electrodes, transitions between any polarized states within the blue region of Figure 2(d) can be achieved with the same level of stability and repeatability. These results exhibit the promising capability of our device for wavefront shaping with well control over reconfigurable arbitrary polarization states.

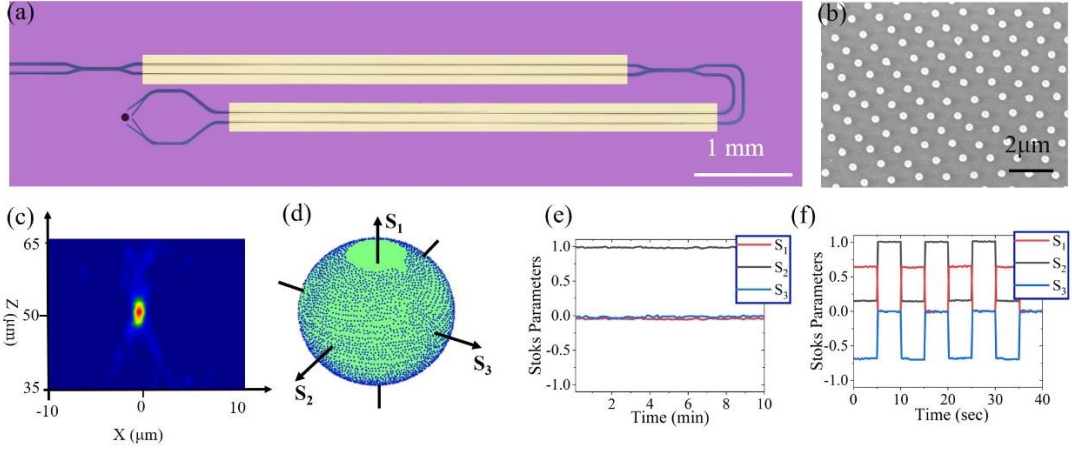

**Figure 2.** (a) The microscope image of the fabricated device. The length of the two sets of electrodes are 0.5mm. (b) SEM image of the fabricated metasurface on the waveguide. (c) The measured intensity profile at X-Z plane. (d) The measured polarization states on the Poincaré sphere of the focal spot. The blue dots on the Poincaré sphere represent the polarization states of the focal spot obtained from experimental testing. (e) Stokes parameters of one of the generated polarization state (45° linear polarization as shown in the inset) as a function of time. (f) The Stokes parameters by switching between two generated polarization states.

In addition to enabling the generation of a single wavefront with reconfigurable polarizations, this scheme has the capability to achieve high-speed switchable multi-functionalities through the incorporation of polarization-dependent metasurfaces, which holds tremendous potential for a wide range of applications. To exhibit this capability, we first introduce a geometric metasurface to the PIC device, enabling the realization of two focal points with orthogonal polarizations at different lateral positions. The metasurface composes of two sets of silicon nanobars as shown in **Fig. 3(a)**, which are designed to realize two focuses based on left circular polarization (LCP) and right circular polarization (RCP) states, respectively. The two sets of nanostructures are spatially multiplexed with a near face centered square unit with the period of the effective wavelength of the $TE_0$ mode in the slab waveguide (SM section 2). In this arrangement, the two sets of nanostructures are positioned in locations with the same local polarization $\psi$. By adjusting the voltages applied to the electrodes, the local polarization $\psi$ can be dynamically switched between LCP and RCP states. Through this control mechanism, the scattered light can be switched between the two designed focal spots, respectively. Meanwhile, the unwanted co-polarization noise will be greatly suppressed due to the phase mismatch[45].

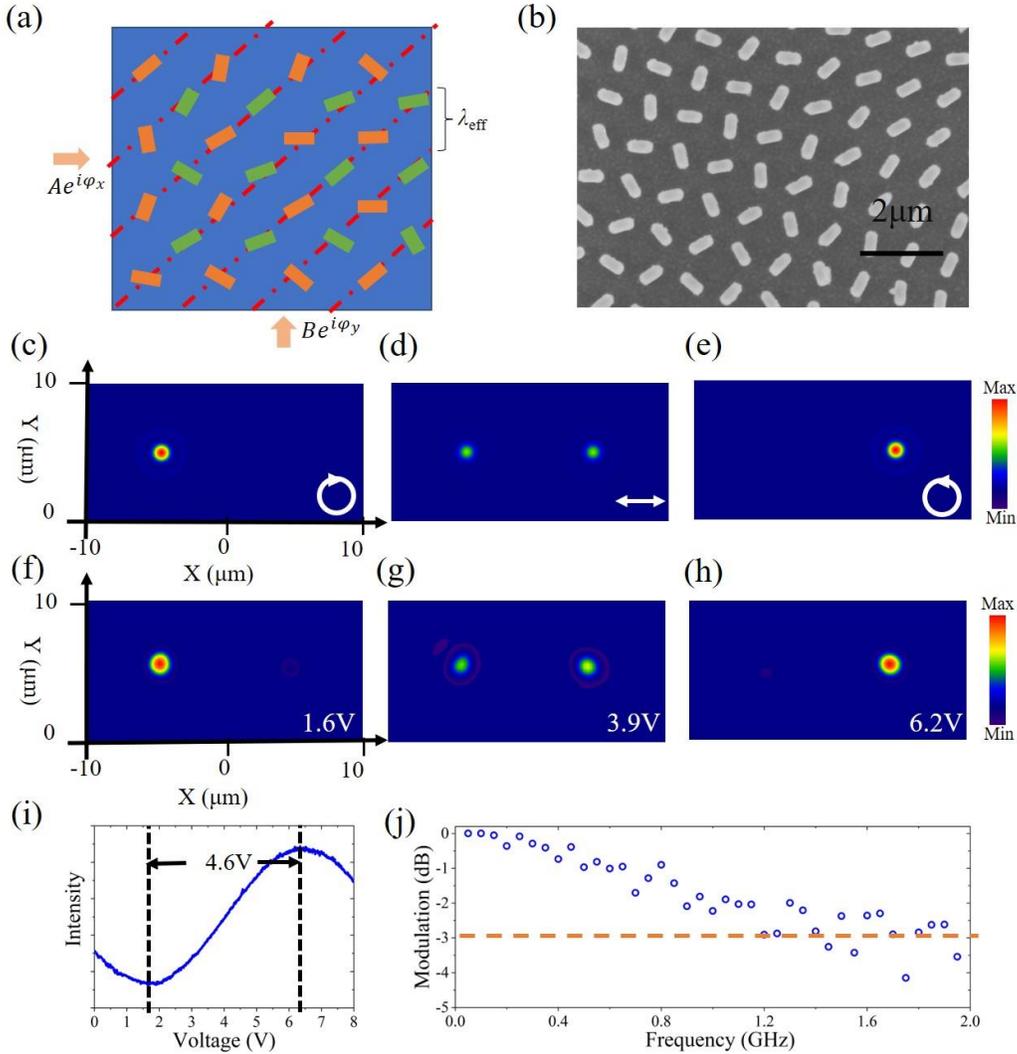

**Figure 3.** (a) The schematic of the switchable metasurface design. (b) SEM image of the fabricated metasurface structure. (c-e) Calculated results of focuses under different local polarization states ψ depicted in the figures. (f-h) The corresponding experimental results. The polarization states are realized by applying different voltages shown in the figures to the electrodes, respectively. (i) The intensity of the right focus in panel f as a function of applied voltage. (j) Peak electro-optic amplitude for modulation frequencies up to 2 GHz.

In the experiment, the silicon nanobars have a uniform length (L) of 300 nm, width (W) of 100 nm, and height (H) of 1000 nm. The corresponding SEM image of fabricated metasurface on the slab waveguide is shown in Fig. 3(b). The two focuses are designed at *x= 5 μm* and *x=-5 μm*, with the center of the metasurface array serving as the origin of coordinates. The focal length is set to $f = 50~\mu m$ for both spots. **Figure 3 (f-h)** show the recorded images of the focuses at different voltages for the phase control. The voltage of the MZI is offset to ensure equal amplitudes in the two arms. By adjusting the voltages applied to the phase shifter, the scattered energy gradually transitions between the two focal spots. The experimental results align well with the calculated images based on the corresponding designed locally polarizations, as shown in Fig. 3(c-e). This result demonstrates the effectiveness of the adopted strategy. A video illustrating the dynamic modulation process is provided in the supplementary material.

To accurately assess the modulation performance, we integrated the recorded focal intensities in Fig. 3 and observed an extinction ratio of approximately 10.2 dB. The electro-optical tunability of the device was evaluated by directing the light from one focus to a high-speed photon detector (FINISAR XPDV21x0(RA)) for analysis (SM section 6). The measured $V_\pi$ of the device is approximately 4.6 V as shown in **Fig. 3(i)** (corresponding to a $V_\pi \cdot L$ of 2.3 V·cm). **Figure 3 (j)** presents the peak electro-optic modulation amplitude for frequencies up to 2 GHz. The result indicates the electro-optic bandwidth of the sample is around 1.4 GHz, providing clear evidence of its gigahertz tunability. Furthermore, the switching speed could be further improved to hundreds of gigahertz by carefully optimized the electrodes and waveguide design.

Apart from the ability to vary the lateral positions of the focal points, this scheme also enables dynamic switching of the focal length, which holds great significance and garners plenty of attentions[51,52]. To demonstrate this concept, two different sets of nanostructures are designed, resulting in two focal points with focal lengths of $f = 35~\mu m$ and $f = 50\mu m$, respectively. **Figure 4(d-f)** shows the recorded images of the focuses by varying the voltage applied to the phase shifter, which agree well to the calculated results (**Fig. 4(a-c)**). The focal points can be dynamically adjusted between these two states or any intermediate states. These integrated lenses with high-speed switchable focal positions and lengths holds promise for future high-speed portable imaging applications, opening up new possibilities in the field.

The high-speed switchable wavefront shaping technique holds immense potential for various applications such as optical communications, imaging, optical computation, sensing, and more. Particularly, the generation of optical orbital angular momentum (OAM) with a large modulation bandwidth is of significant importance in the realm of optical communication[53]. We replace the focus beams with two OAMs with different topological numbers and locations by mapping the required phase distributions to the two sets of metasurfaces. Fig. 4(j-l) shows the typical recorded OAM images by varying the voltages applied to the electrode of the phase shifter, which agree well with the calculated results under certain corresponding polarizations shown in Fig. 4(g-i). The left OAM beam is designed with l=+1 in LCP state and the right one is designed with l=-1 in RCP state. The results exhibit that the integrated device has the capability to dynamically manipulate OAM beams with various topological numbers, offering promising applications in optical communications and manipulations. These findings indicate the feasibility of utilizing a geometric phase-based spatial multiplexing multi-channel device within our platform.

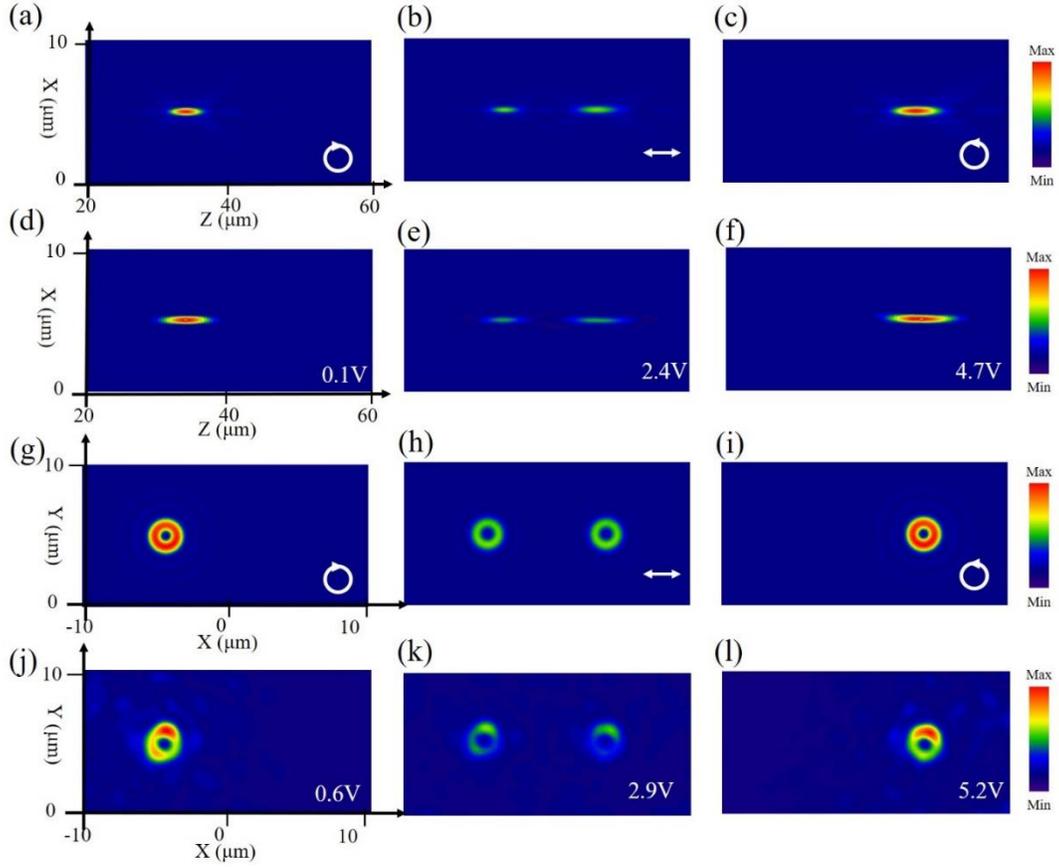

**Figure 4.** (a-c) Calculated results of switchable focal length under different local polarization states depicted in the figures. (d-f) The corresponding experimental results realized by applying different voltages to the electrodes as shown in the figures, respectively. (g-i) Calculated results of switchable OAM beams with switchable topological charges under different local polarization states depicted in the figures. (j-l) The corresponding experimental results realized by applying different voltages to the electrodes as shown in the figures, respectively.

The aforementioned multi-function switchable devices are designed with a geometric phase constraint limited to circular polarizations. Indeed, it is important to note that this constraint can be extended to encompass any arbitrary orthogonal polarization states by considering the geometric phase and propagation phase of birefringent nanostructures for free-space light[54]. It holds significant importance in applications involving polarization optics, offering enhanced versatility and adaptability. However, this scheme is still awaiting clarification in the guided-wave-driven metasurfaces due to the overlapping of in-plane waveguide mode and the nanostructures. Meanwhile, the scattering efficiency of the nanostructures on the waveguide exhibits a strong dependence on their size, and this behavior differs significantly from the transmission characteristics of metasurfaces in the free-space configuration. Here, we present the successful engineering and dynamic switching of wavefront in arbitrary polarization states with the metasurface on PIC device. When a birefringent nanostructure scatters the waveguide mode with an arbitrary local polarization of $\Psi_1$, the scattered light can be decomposed into a pair of orthogonal polarization states $\Psi_2^+$ and $\Psi_2^-$, respectively. The two decomposed components can be expressed as:

$$O^{\pm}e^{i\phi^{\pm}} = \hat{\xi}_2^{\pm} \begin{bmatrix} T_x & T_{xy} \\ T_{xy} & T_y \end{bmatrix} \Psi_1 \qquad (4)$$

Where $\hat{\xi}_2^{\pm}$ represent the projection operators for the polarization states of $\Psi_2^+$ and $\Psi_2^-$, respectively. $O^{\pm}$ and $\varphi^{\pm}$ are the amplitude and phase of the scattered light after the projection operation, respectively. Considering the physical characteristics of linear birefringent metasurfaces and their ability to consistently convert input polarization states into output polarization states with arbitrary phase delays, it is necessary for the input and output polarization states to be a pair of orthogonal polarization states. On the other hand, the required Jones matrix $\begin{bmatrix} T_x & T_{xy} \\ T_{xy} & T_y \end{bmatrix}$ could be implemented by utilizing rectangular nanostructures and adjusting their dimensions to impose arbitrary phase delays upon two orthogonal polarization states. The aforementioned outcome can be comprehended as the amalgamation of both the propagation and geometric phases within a singular component. By simultaneously adjusting the size and orientation of the birefringent nanostructure, it becomes possible to impart desired phases onto the orthogonal polarization states. To design the nanostructure with desired Jones matrix, we built a phase-amplitude response library with regard to the rectangular nanostructure dimensions on slab waveguide using full-wave simulations (SM section 5). Here, the width and the length of the nanostructure are varied between 100 nm to 500 nm and the height is 1000 nm as above structures. The amplitude and the phase of the scattering upon different dimensions are shown in **Fig. 5 (a-b)**. A genetic algorithm is employed to retrieve the required dimensions for the desired Jones matrix with near uniform scattering intensity. Finally, arbitrary phase profiles in the independent polarization channels could be realized by spatially arranging the retrieved nanostructures.

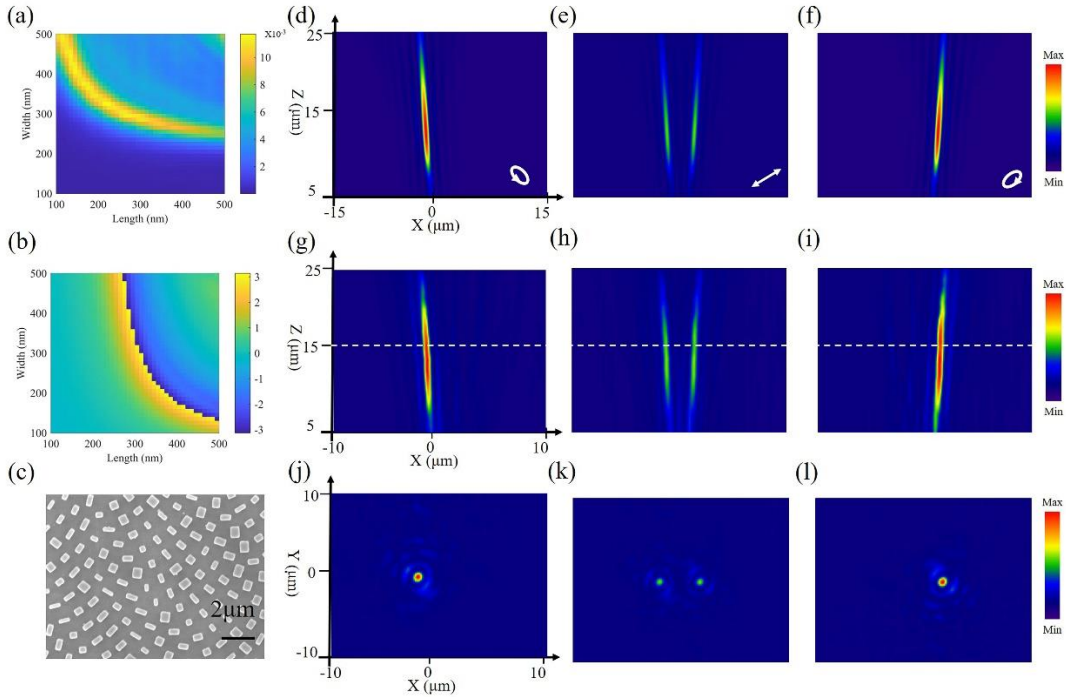

**Figure 5.** (a) The simulated intensity and (b) phase of the scattered light when the rectangular nanostructure's length and width varies from 100 to 500 nm. (c) SEM image of the fabricated metasurface for Bessel beams. (d-f) Calculated results of switchable Bessel beams under different local polarization states depicted in the figures. (g-i) The corresponding experimental results by adjusting voltages applied to the electrodes, respectively. (j-l) The corresponding cross-section images along the white dashed lines in panel g-i, respectively.

Similar to the geometric phase based metasurfaces demonstrated above, by varying the polarization distribution within the planar waveguide region, we can switch the wavefronts designed upon the arbitrary independent polarization channels. As an example, here we demonstrate an integrated generation of switchable Bessel beams with arbitrary polarizations, which have garnered significant interest across a wide range of research fields. The polarization states of the two Bessel beams are designed with $1/\sqrt{5}[2\ \ e^{i\pi/4}]^T$ and $1/\sqrt{5}[1\ \ 2e^{5i\pi/4}]^T$, respectively. Figure 5(c) depicts the SEM images of fabricated sample. **Figure 5 (g-i)** illustrate the featured images through the adjustment of voltages applied to the MZI and the phase shifter. It is clearly shown the formation of Bessel beams in different directions (±5°) of the scattered light, these adjustments align with the calculation results displayed in **Fig. 5 (d-f)**, considering the corresponding polarizations. **Figure 5 (j-l)** shows the recorded lateral images along the white dash lines in **Fig. 5 (g-i)**, respectively. The integrated generation of switchable non-diffracting Bessel beams with arbitrary polarization states demonstrates the feasibility of our method for achieving switchable functionalities in any arbitrary polarization state. Moreover, it introduces exciting new prospects for applications involving special beam characteristics.

We have demonstrated the capability of wavefront shaping and switching upon any orthogonal polarization states. Indeed, this high-speed switchable scheme could also be extended to nonorthogonal polarization states as well. By incorporating complex unit cell, interleave design, as well as artificial intelligence, the traditional limitation of two independent orthogonal channel could be broken, enabling the realization of more polarization-determined and switchable channels[9]. This strategy not only enables the generation of specialized optical beams but also facilitates the implementation of various switchable and complex functionalities. In addition to generate switchable wavefronts in uniform polarization states, the PIC-driven metasurface exhibits the potential to produce complex vector wavefronts. Furthermore, this approach can also utilize alternative high-speed modulation PIC platforms and mechanisms, in addition to LNOI, such as carrier depletion, enabling increased levels of integration and compatibility with CMOS technology. The out-of-plane extraction efficiency of the integrated metasurface in our device can reach to 11% with numerical simulation, which can be further improved by optimizing the geometric parameters of the metasurfaces and the waveguide[50,55] (SM, section 8). These factors expand the applicability across a wide range of fields and applications.

In summary, we have proposed and demonstrated an integrated electro-optical platform with PIC-driven metasurface on LNOI. By integrating an electrically controlled Mach-Zehnder interferometer (MZI) and a phase shifter, we construct a focusing beam with high-speed adjustable polarizations across almost the entire surface of the Poincaré sphere. Based on the reconfigurable polarizations across the waveguide, switchable focusing beams with lateral focal positions and focal lengths, OAM beams and Bessel beams are demonstrated. Our approach opens up possibilities for achieving more switchable functionalities with complex polarizations. The modulation bandwidth is measured to be 1.4 GHz and it is potential to be improved to hundreds of gigahertz based on the electro-optical effect of LN. The demonstrated high-speed switchable PIC-driven metasurface provides a promising platform for a variety of applications in high-capacity optical communication, fast optical computation, imaging and sensing. This integrated platform offers significant opportunities for advancing these fields and enabling a wide range of practical applications.

*Code and Data Availability*

Structural parameters, simulated and experimental data have been provided within the main text and Supplementary Material of this paper. All the other data that support the findings of this study are available from the corresponding authors upon reasonable request.

*Acknowledgments*

This work was supported by the National Key R&D Program of China 2019YFA0705000, National Natural Science Foundation of China (grant Nos. 12192251, 12274134, 12174186，

# Supplementary material for "Gigahertz-rate-switchable wavefront shaping through integration of metasurfaces with photonic integrated circuit"

**Haozong Zhong,**[a, †] **Yong Zheng,**[a, †] **Jiacheng Sun,**[b, †] **Zhizhang Wang,**[b] **Rongbo Wu,**[a] **Ling-en Zhang,**[a] **Youting Liang,**[a] **Qinyi Hua,**[a] **Minghao Ning,**[a] **Jitao Ji,**[b] **Bin Fang,**[c] **Lin Li,**[a, e, *] **Tao Li,**[b, *] **Ya Cheng,**[a, d, e, *] **Shining Zhu**[b, *]

[a]State Key Laboratory of Precision Spectroscopy, School of Physics and Electronic Science, East China Normal University, Shanghai 200241, China
[b]National Laboratory of Solid State Microstructures, College of Engineering and Applied Sciences, Nanjing University, Nanjing, 210093, China
[c]College of Optical and Electronic Technology, China Jiliang University, Hangzhou 310018, China
[d]State Key Laboratory of High Field Laser Physics and CAS Center for Excellence in Ultra-Intense Laser Science, Shanghai Institute of Optics and Fine Mechanics (SIOM), Chinese Academy of Sciences (CAS), Shanghai 201800, China
[e]Collaborative Innovation Center of Extreme Optics, Shanxi University, Taiyuan 030006, China

*Lin Li, lli@lps.ecnu.edu.cn; Tao Li, taoli@nju.edu.cn; Ya Cheng, ya.cheng@siom.ac.cn; Shining Zhu, zhusn@nju.edu.cn

## S1: The amplitude and phase evolution in the slab waveguide

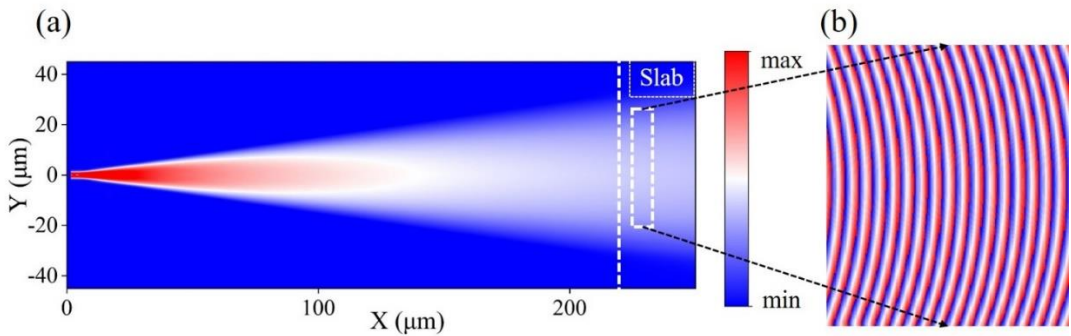

**Figure S1.** (a) Simulated amplitude evolution associated with the guided mode in the taper and slab waveguide. (b) The phase evolution in slab waveguide.

Here, we present the variations in amplitude and phase of waveguide mode during its propagation from an adiabatic taper to a planar waveguide. Figure S1(a) illustrates the process, where light travels from a single mode ridge waveguide to a slab waveguide through an adiabatic taper of 200 μm. The mode size gradually increases to 60 μm, further enlarging within the slab waveguide, resulting in a region of uniform amplitude. In this case, the width at the end of taper is 60 μm. By designing the length and width of the taper, the area of uniform amplitude can be further optimized. Figure 1(b) illustrates the phase evolution of light within the slab waveguide, presenting an arc-shaped phase profile. Fitting results indicate that the center of the arc corresponds to the starting point of the taper. Although the arcs have very large radii of curvature, they can be approximated as near planar waves within a limited area. Based on the simulation result, the precise polarization states distribution, which formed by the superposition of the two orthogonally propagated waveguide modes, could be retrieved. The functionalities in the main text are also designed based on the simulated amplitude and phase evolution, taking into account the non-planar nature of the waveguide modes.

The slab waveguide mode will travel a relatively long distance within the device before it reaches the boundary of the waveguide after passing through the metasurface. Meanwhile, the waveguide mode in the slab gradually expands after passing through the tapers as illustrated in Figure S1. (a). Consequently, following the long-distance propagation, when the light encounters the boundary and is reflected back, the intensity of the guided mode light will be significantly attenuated. In addition, the design of the metasurface relies on the wavefront and the propagation direction of the waveguide mode, while the reflected light dose not consistent with them. As a result, the influence of the reflected light on the performance of the device can be safely disregarded.

## S2: Design of polarization dependent guided-wave-driven metasurface

Once we obtain the amplitude and phase evolution in the planar waveguide for the two orthogonally propagated channels, we can extract the actual distribution of identical polarizations $\psi$ on the waveguide. These distributions appear as a series of lines, similar to the dashed lines shown in Fig. 3(a) in the main text. The polarization state can be dynamically modulated to any desired state by adjusting the voltages applied to the electrodes of the MZI and the phase shifter.

Without loss of generality, here we take LCP state as the reference design polarization as an example. By considering the phase evolution, we can also identify a series of arcs with the same phase (as shown in Fig. S1(b)). These arcs, together with the lines of identical polarization, form an array of crosses, where the metasurface nanostructures will be located. To achieve the desired beam (beam 1), the metasurfaces are designed to fulfill the corresponding phase and amplitude distributions based on the LCP state. Simultaneously, another array of crosses can be obtained by introducing a phase difference of $\pi$ to the previous series of arcs, and aligning them with the lines of identical polarization. We can design a separate metasurface to generate a different beam (beam 2) based on the right circularly polarized (RCP) state.

When the polarization of the lines is adjusted to be LCP, only beam 1 can be formed, while only beam 2 can be formed when the polarization is adjusted to be RCP. It is worth noting that the choice of the reference design polarization can be arbitrary, as discussed in the main text.

## S3: The polarization extinction ratio of the generated beam

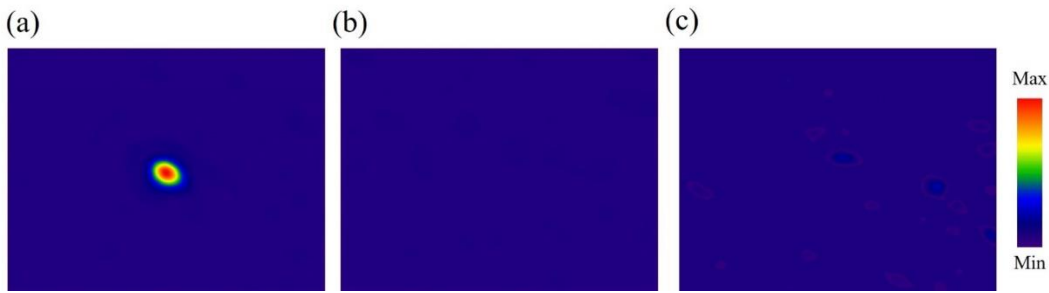

**Figure S2. (a)** The cross-section image of the focal spot in Figure 2(c) at *X-Y* plane. **(b)** The corresponding image after performing polarization analyzing at the orthogonal polarization state. **(c)** The recorded image with increased exposure time by twenty times.

We performed measurements to determine the polarization extinction ratio (PER) of the focal point. In Fig. S2(a), we show a cross-sectional image of the focal spot in Fig. 2(c) in the main text at the X-Y plane. Fig. S2(b) displays the corresponding image after polarization analysis at the orthogonal state. It can be observed that the focal spot disappears after polarization filtering. To retrieve the energy of the spot under this condition, we increased the exposure time of the camera by a factor of twenty. The resulting image is presented in Fig. S2(c), where the focal spot is not visible, indicating a high degree of polarization purity. Let P1 and P2 represent the integrated intensities of the two focal positions in Figure S2(a) and Figure S2(c), respectively. The PER of the focal spot can be calculated using the formula $10 * \log_{10}( (20 * P1) /P2)$. The calculated value is 20.79 dB, indicating a minimum PER of 20 dB for the focal spot.

## S4 Calculation of the propagation process of scattered light

This section introduces the calculation of the propagation of electromagnetic waves using the Huygens-Fresnel principle. According to this principle, each point on a wavefront can be regarded as a secondary wave source. By superimposing the amplitudes and phases of these secondary wave sources, the wavefront at the next moment can be obtained. In the calculation, each metasurface structure is treated as a point source. The initial amplitudes and phases of these point sources are determined by considering the simulated data of individual nanostructures and the propagation conditions of the waveguide mode. Based on the principles of the Huygens-Fresnel principle, the distribution of the light field in surrounding space can be derived with these point sources.

## S5: Simulation of individual element of guided-wave-driven metasurface

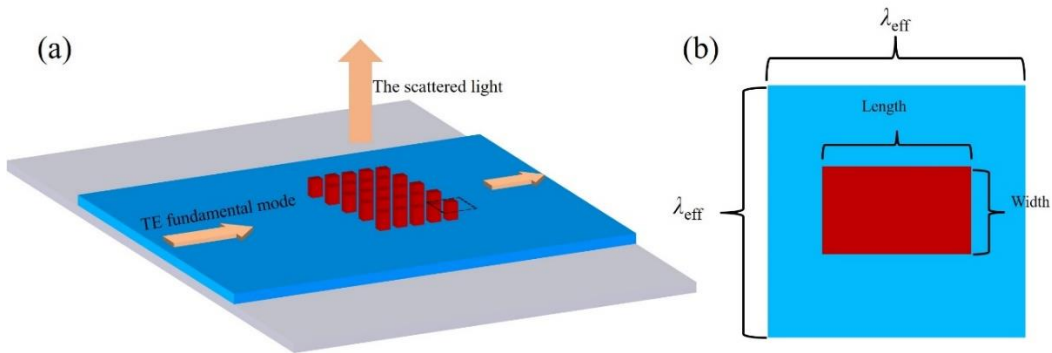

**Figure S3.** (a) Simulation schematic of metasurface structure of different sizes. (b) Schematic of a unit cell of the metasurface.

Different from the simulation method for the unit cell of metasurface in free space, the periodic boundary condition cannot be directly applied to the guided-wave-driven metasurface. In order to accurately evaluate the scattering performance of the nanostructure, we employed a 5x5 metasurface array arranged on a waveguide for convenient simulation. It is observed that the amplitude and phase of the scattered light do not exhibit significant variations in larger arrays. The period of the array is chosen to match the effective wavelength of the waveguide mode, allowing for near-plane scattering. The waveguide width is set to 10μm to create a mode spot area large enough to accommodate the TE fundamental mode. This ensures that the entire array operates within a region of uniform amplitude. By varying the length and width of the nanostructures within the range of 100nm to 500nm, we obtained the intensity and phase of the scattered light associated with the corresponding nanostructures interacting with the waveguide mode.

## S6: Optical measurement setups

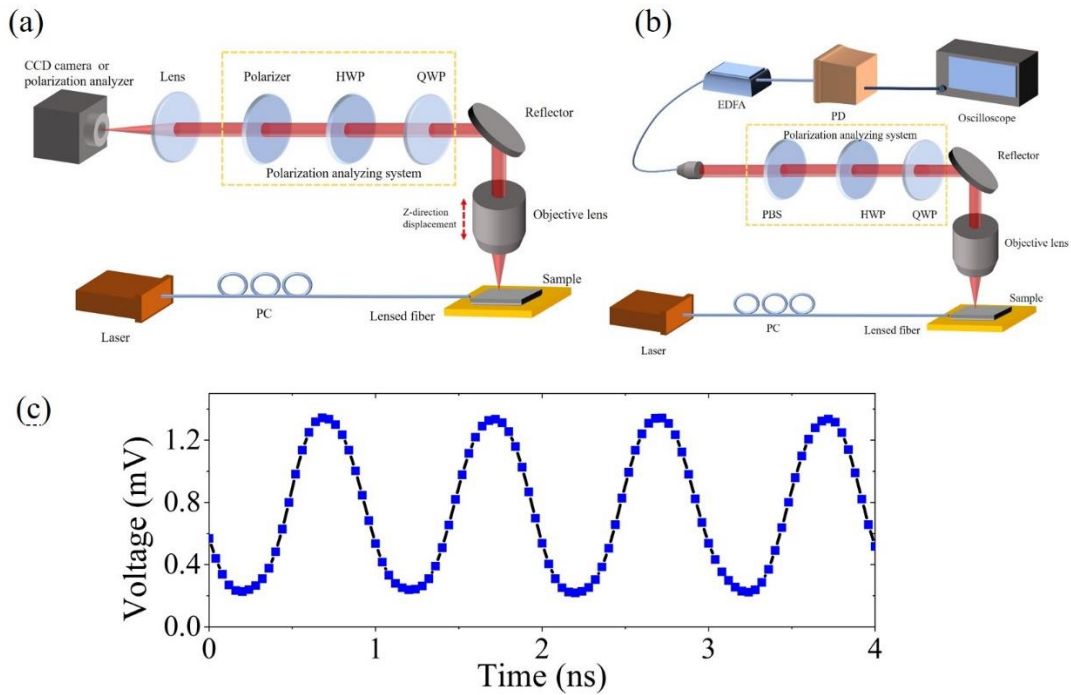

**Figure S4.** (a) The optical measurement setup to image the scattered light from the device. (b) The optical measurement system to characterize the modulation performance. The optical components within the yellow dashed box represent the polarization analysis system, which can be removed when polarization analysis is not required. (c) The test result under modulation by a 1 GHz sinusoidal voltage signal.

Figure S4(a) illustrates the measurement setup, where a 1550 nm laser passes through a polarization controller and is coupled to the waveguide on the sample via a single-mode fiber lens. The scattered light from the metasurface is collected by an objective lens and then imaged with a CCD camera. The objective lens is mounted on a high-precision translation stage for z-direction displacement. By gradually adjusting the translation stage along z-direction, the images are captured at different heights and subsequently overlapped to reconstruct a three-dimensional stereoscopic image.

Meanwhile, this setup can be utilized for polarization state characterization. For this purpose, an polarization analyzing system is employed before the CCD camera. To characterize the of polarization reconfigurable performance of the device at high speed, the CCD camera is replaced with a high-speed Avalanche Photodiode (Thorlabs, APD430C/M, DC - 400 MHz). By adjusting the quarter waveplate and polarizer, precise derivation of the Stokes parameters is achieved [1]. In addition, the polarization state is characterized using a commercial polarization analyzer (PA) (Thorlabs PAX1000IR2/M) for lower modulation speeds and longer measurement durations.

Figure S4(b) illustrates the experimental setup employed for characterizing the high-speed modulation performances. The light emitted by the metasurface is coupled into a fiber and amplified using an Erbium-Doped Fiber Amplifier (EDFA). Subsequently, the light is detected by a high-speed photon detector (FINISAR XPDV21x0(RA)). Finally, the optoelectronic detector's signal is collected by an oscilloscope (Tektronix MSO64B). Figure S4(c) illustrates the test result under modulation by a 1 GHz sinusoidal voltage signal. To examine the polarization properties of the modulated light, a set of polarization analyzing systems can be added before the coupling system, as depicted in the figure.

## S7: Analysis of the Performance of the Polarization State Reconfiguration

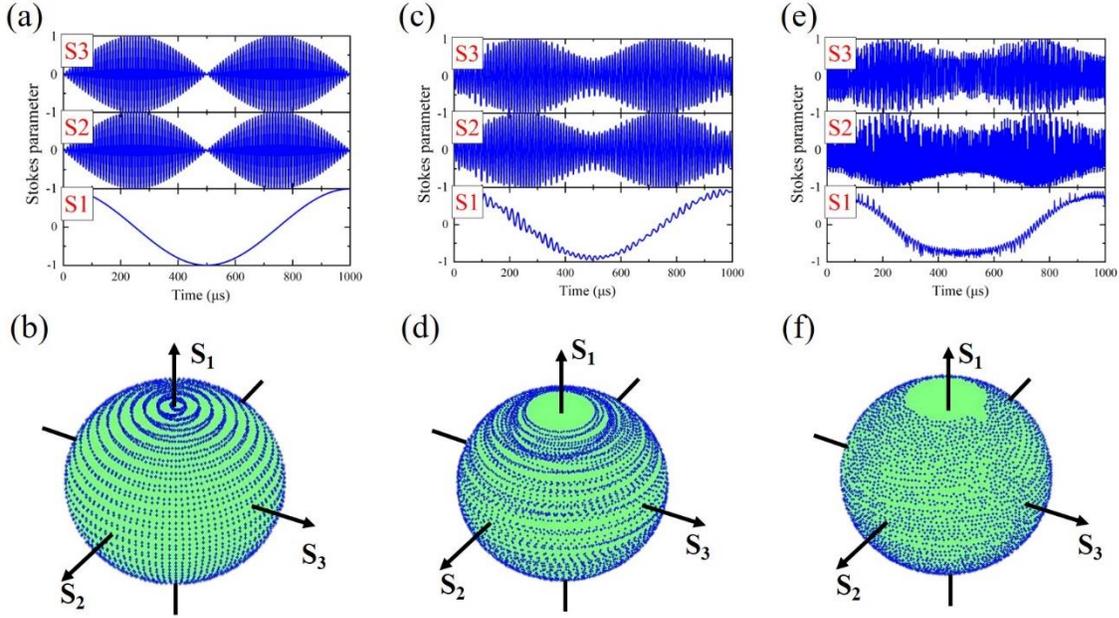

**Figure S5.** (a) Theoretical analysis of time-varying Stokes parameters when applying triangle wave signals with frequencies of 1 kHz and 100 kHz to the electro-optic phase shifters, respectively, and (b) the corresponding polarization states' distributions on the Poincaré sphere. (c) Theoretical analysis of time-varying Stokes parameters considering uneven splitting ratio in the directional coupler, and (d) the corresponding polarization states' distributions on the Poincaré sphere. (e) The measured Stokes parameters as a function of time when applying the two triangle wave signals to the phase shifters, respectively, and (f) the corresponding polarization states' distributions on the Poincaré sphere. Here the blue dots represent the polarization states of the focal spot.

In order to assess the performance of the polarization reconfiguring with our device, two triangle wave signals with frequencies of 1 KHz and 100 KHz are applied to the electrodes of the Mach-Zehnder interferometer (MZI) and phase shifter, respectively. In this scenario, the optical electrical field of the two guided waves can be expressed as follows [2]:

$$\begin{pmatrix} |a\rangle \\ |b\rangle \end{pmatrix} = C_\varphi \times C_{\text{DC}} \times C_\theta \times C_{\text{DC}} \times \begin{pmatrix} |1\rangle \\ 0 \end{pmatrix} \tag{S1}$$

where $C_{DC}$, $C_\theta$ and $C_\varphi$ are the transfer matrices of directional couplers, first and second phase shifters, respectively. $|1\rangle$ is the input optical electrical field. $|a\rangle$ and $|b\rangle$ represent optical electrical field of the two guided waves at the tapers, corresponding to the relevant portion of Equation (1) in the main text:

$$|a\rangle + |b\rangle = \begin{bmatrix} Ae^{i\varphi_x} \\ Be^{i\varphi_y} \end{bmatrix} \quad (S2)$$

Figure S5. (a) shows theoretical analysis of time-varying Stokes parameters when applying triangle wave signals with frequencies of 1 kHz and 100 kHz to electro-optic phase shifters, with ideal 1:1 splitting ratio in the directional couplers. However, the splitting ratio of the directional couplers is difficult to reach 1:1 due to the fabrication imperfections. Additionally, the waveguides here support higher order modes, which could be excited with the fiber lenses in the waveguide mode coupling process. These higher order modes will impact the reconfigurated polarizations of the focal spot, though their intensities are much weaker than the fundamental mode. Figure S5. (c, d) present the theoretically calculated results after taking these factors into account. Here, we assume a splitting ratio of 5:4 between the two ports of the directional coupler, and a ratio of 10:1 between the fundamental mode and higher order modes. The results show that these two factors significantly impact the Stokes parameters, and the polarization states cannot fully cover the surface of the Poincaré sphere. Figure S5. (e, f) displays the experimentally measured Stokes parameters and the corresponding reconfigured polarization states of the focus on the Poincaré sphere. The distribution of the polarization states agrees with the theoretical analysis in Figure S5. (c,d). Further development will be addressed in the future work to improve the performance. An additional electrically controlled interferometer can be introduced to improve the splitting ratio and make the reconfigured polarization state fully cover the Poincaré sphere. Meanwhile, stratagems to filter out high-order modes can be involved to reduce the impact of high order modes.

## S8: The efficiency of the device

The out-of-plane extraction efficiency of the integrated metasurface is quantified as the ratio of the optical power carried by the generated beams to the optical power within the input guided waves. Under these prescribed conditions, we conducted numerical simulations to the metasurfaces depicted in Figures 2. (b), Figures 3. (b) and Figures 5. (c), to assess their efficiency. The results indicate efficiencies of 0.2%, 6%, and 11% respectively. In experimental endeavors, the overall efficiency is contingent upon the coupling efficiency between the fiber lens and the waveguide mode, the propagation loss of LN waveguide, and the out-of-plane extraction efficiency of the metasurface. Specifically, the coupling efficiency between the fiber lens and the waveguide mode was quantified at about -7dB. The propagation loss within the LN waveguide is of a magnitude low enough to be negligibly significant. As a result, the overall efficiency of the devices is estimated to be approximately 0.04%，1.2% and 2.2%[5,6]. The overall efficiency can be further increased by optimizing the geometric parameters of the metasurfaces, the underneath waveguide and the coupling method between the fiber mode and the waveguide mode.

## S9: Fabrication methods

Our device is fabricated on a 4-inch X-cut LNOI wafer. It is composed of a 500 nm-thick LN layer bonded to a buried silica ($SiO_2$) layer on a silicon support substrate with a thickness of 500 μm. The fabrication process involved the utilization of Photolithography Assisted Chemo-Mechanical Etching (PLACE) technology for photonic integrated circuits (PICs) [3].

The fabrication process of PIC comprises four main steps. Firstly, a 600 nm-thick chromium (Cr) film is deposited onto the surface of the LNOI substrate using magnetron sputtering. Subsequently, the Cr film on the LNOI sample is patterned into the waveguide mask utilizing a space-selective femtosecond laser system (PHAROS, LIGHT CONVERSION Inc.). Following that, a chemo-mechanical polishing (CMP) process is employed to selectively etch the thin film of lithium niobate using a wafer polishing machine (UNIPOL802, Kejing Inc.). In this step, the lithium niobate thin film, protected by the Cr mask, remains intact after the CMP process and served as the waveguides. Finally, the fabricated structure is immersed in a chromium etching solution to remove the Cr mask.

After completing the LN waveguide fabrication, two layers of LOR5B film and AZ5214 film are spin-coated on the sample. Subsequently, an ultraviolet (UV) exposure is conducted to define the electrode patterns. After the development process, a 300 nm-thick gold film is deposited onto the wafer. Then, a lift-off process is employed to transfer the patterns.

Following the electrode fabrication, a layer of α-Si is deposited on the chip using plasma-enhanced chemical vapor deposition (PECVD). Subsequently, the metasurface patterns are written on an electron beam resist (PMMA) through electron beam lithography (EBL). After development, a lift-off process is employed to define the chromium pattern and served as a hard mask for the following dry etching process. The α-Si pattern is transferred afterwards by dry etching process with a mixture of C4F8 and SF6 plasma (HSE200, Naura) and a remove of Cr hard mask [4].